\documentclass[A4,prl, superscriptaddress ,preprintnumbers,amsmath,amssymb,floatfix,footinbib,twocolumn]{revtex4-1}
\usepackage{graphicx}% Include figure files
\usepackage{epstopdf}
\usepackage{bm}% bold math
\usepackage{subfigure}
\usepackage{amsmath}
\usepackage{hyperref}
\usepackage{color}
\usepackage{multirow}
\usepackage{lpic}

\newcommand{\eqn}[1]{Eq.~\ref{#1}}
\newcommand{\fig}[1]{Fig.~\ref{#1}}

\newcommand{\kB}{k_\textrm{B}}

\newcommand{\LW}{\Gamma_L}
\newcommand{\DW}{\Gamma_D}
\newcommand{\DWfit}{\DW^{\textrm{fit}}}
\newcommand{\DWprt}{\DW^{\textrm{PRT}}}
\newcommand{\Isat}{I_\textrm{sat}}
\newcommand{\HFS}{f_\textrm{HFS}}

\begin{document}

\title{
Quantum-Limited  Spectroscopy
}
\author{G.-W.\ Truong}%\email[]{} 
\affiliation{School of Physics, The University of Western Australia, Perth, Western Australia 6009, Australia}
\affiliation{Institute for Photonics and Advanced Sensing (IPAS) and School of Chemistry and Physics, The University of Adelaide, Adelaide, SA 5005, Australia}
\author{J.\ D.\ Anstie}%\email[]{} 
\affiliation{Institute for Photonics and Advanced Sensing (IPAS) and School of Chemistry and Physics, The University of Adelaide, Adelaide, SA 5005, Australia}
\affiliation{School of Physics, The University of Western Australia, Perth, Western Australia 6009, Australia}
\author{E.\ F.\ May}%\email[]{} 
\affiliation{Centre for Energy, School of Mechanical and Chemical Engineering, The University of Western Australia, Crawley, Western Australia 6009, Australia}
\author{T.\ M.\ Stace}\email[]{theoretical questions: stace@physics.uq.edu.au} 
\affiliation{ARC Centre for Engineered Quantum Systems, University of Queensland, Brisbane 4072, Australia}
\author{A.\ N.\ Luiten}\email[]{experimental questions: andre.luiten@adelaide.edu.au}  
\affiliation{Institute for Photonics and Advanced Sensing (IPAS) and School of Chemistry and Physics, The University of Adelaide, Adelaide, SA 5005, Australia}
\affiliation{School of Physics, The University of Western Australia, Perth, Western Australia 6009, Australia}

\date{\today}

\begin{abstract}

Spectroscopy has an illustrious history delivering serendipitous discoveries and providing a stringent testbed for new physical predictions, including applications from trace materials detection, to understanding the atmospheres of stars and planets, and even constraining cosmological models. Reaching fundamental-noise limits permits optimal extraction of spectroscopic information from an absorption measurement.  Here we demonstrate a quantum-limited spectrometer that delivers high-precision  measurements of the absorption lineshape. These measurements yield a ten-fold improvement in the accuracy of the excited-state (6P$_{1/2}$) hyperfine splitting in Cs, and reveals a breakdown in the well-known Voigt spectral profile. We develop a  theoretical model that accounts for this breakdown, explaining the observations to within the shot-noise limit. Our model enables us to infer the thermal velocity-dispersion of the Cs vapour with an uncertainty of 35ppm within an hour. This allows us to determine a value for Boltzmann's constant with a precision of 6ppm, and an uncertainty of 71ppm.

%Spectroscopy has an illustrious history delivering   serendipitous discoveries and providing a stringent testbed for new physical predictions.  It  has  a broad range of applications from trace materials  detection, to understanding the atmospheres of stars and planets and even constraining  cosmological models.  To  extract the maximum    information from an absorption measurement, it is vital to  explore the limits of spectroscopic accuracy and precision.  Here we demonstrate a quantum-limited atomic spectrometer that delivers high-precision  measurements of the absorption line-shape.  These measurements yield a ten-fold improvement in the accuracy of the excited-state (6P$_{1/2}$) hyperfine splitting in Cs, and reveal a significant breakdown in the well-known Voigt  spectral profile.We develop a  theoretical model that accounts for this breakdown, and  explains   the observations   to within the shot-noise limit. Our model enables us to infer the thermal velocity-dispersion of the Cs vapour with an uncertainty of 35 ppm in under one hour. This, in turn, allows us to determine a value for Boltzmann's constant with a precision of 6 parts-per-million (ppm), and an uncertainty of 71 ppm.

\end{abstract}

\pacs{
}

\maketitle
Spectroscopy is a vital tool for both fundamental and applied studies. It is critical to  drive improvements in both precision and accuracy in order to gain maximum physical information about the quantum absorbers -- even as they are perturbed by the probe. The immediate motivation for this work has arisen out of a call to the scientific community to develop new techniques to re-measure Boltzmann's constant, $\kB$, in preparation for a redefinition of the kelvin \cite{fellmuth2006}; however, advances in absorption lineshape measurement and theory will find applications in accurate gas detection and monitoring  \cite{Stewart2011}, studies of planetary atmospheres \cite{drossart2005infrared},  thermometry  in tokamaks \cite{koch2012} and   understanding distant astrophysical processes \cite{grefenstette2014asymmetries,hayato2010}. The accurate measurement of the natural linewidth and transition frequencies, which can be directly related to the atomic lifetime, level structure and transition probabilities, is important   for testing atomic physics \cite{Wood21031997,jonsson2996,brage1994}.

Here we present measurements of a transmission lineshape of cesium (Cs) vapor with a quantum-limited  transmission uncertainty of 2\,ppm in a 1 second measurement, which is to our knowledge, a factor of 16 times superior to anything previously demonstrated \cite{moretti2013}. This extreme precision allows us to directly detect subtle lineshape perturbations that have not been previously observed.  This observation prompted  the development of a theoretical model, which now allows us to discriminate between the internal atomic state dynamics and their external motional degrees.  Using the model we are able to  estimate the velocity dispersion of the atoms with a precision (standard error) of 53\,ppm  during a single line scan, taking $\sim 30$ seconds.  This is consistent with the the sample standard deviation (also 53 ppm) over multiple  scans, demonstrating the excellent reproducibility of the spectrometer.  The measurement precision averages down to  3.7\,ppm after 200 scans.  These values are $\sim7$ times better than the previous best results~\cite{lemarchand2010,djerroud2009,moretti2013}, and also yield a 10-fold improvement  in the uncertainty of the excited state hyperfine splitting in Cs~\cite{steck2003cdl,udem1999}. Our measurement of the homogenous broadening component of the line shape has a precision within a factor of two of the best ever measurement of natural linewidth.  Modest improvements in the probe laser performance would deliver a new result for the excited state lifetime of Cs in a system that is experimentally and theoretically much simpler than that typically used for lifetime studies \cite{rafac1999,PhysRevLett.91.153001}.

\begin{figure*}
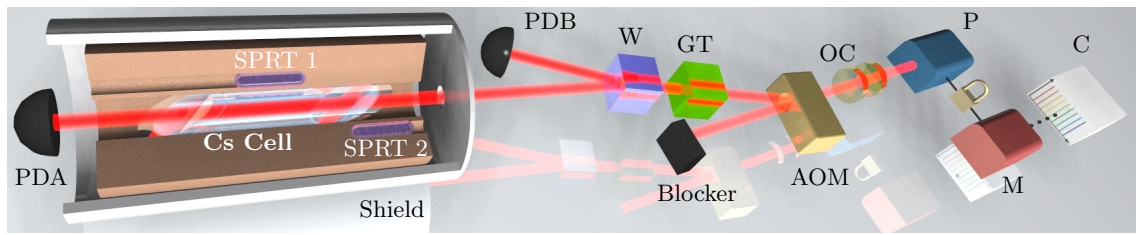

\begin{center}

 %Lit appearance in laser; two SPRT
\begin{lpic}[]{schematic11(15cm)}				
	\lbl[lb]{122,26;$\textcolor{black}{\textrm{{P}}} $}
	\lbl[lb]{127,5;$\textcolor{black}{\textrm{{M}}} $}
	\lbl[lb]{136,23;$\textcolor{black}{\textrm{{C}}} $}
	%\lbl[lb]{110,0;Frequency Measurement}
	\lbl[t]{106,24;$\textcolor{black}{\textrm{{OC}}} $}
	\lbl[lb]{100,6;$\textcolor{black}{\textrm{{AOM}}} $}
	\lbl[lb]{83,4;$\textcolor{black}{\textrm{{Blocker}}} $}
	\lbl[bl]{85.5,23;$\textcolor{black}{\textrm{{GT}}} $}
	\lbl[lb]{78,24;$\textcolor{black}{\textrm{{W}}} $}
	\lbl[lb]{25,10.5;$\textbf{\textcolor{white}{\textrm{{Cs Cell}}}} $}
	\lbl[lb]{29,21;${\textcolor{white}{\textrm{{SPRT 1}}}} $}
	\lbl[lb]{43,9.8;${\textcolor{white}{\textrm{{SPRT 2}}}} $}
%	\lbl[lb]{32,21;$\textbf{\textcolor{white}{\textrm{{PRT}}}} $}
%	\lbl[lb]{30,17.5;$\textcolor{white}{\textrm{{Cs cell}}} $}
%	\lbl[lb]{32,23;$\textcolor{white}{{\textrm{PRT}}} $}
	\lbl[lb]{45,2;$\textcolor{black}{\textrm{{Shield}}} $}
	\lbl[lb]{1,6;$\textcolor{black}{\textrm{{PDA}}} $}
	\lbl[lb]{66,26;$\textcolor{black}{\textrm{{PDB}}} $}
\end{lpic}%remove "draft" to get rid of grid lines
\caption{Schematic of a high-accuracy linear absorption spectrometer based on a probe laser (P) that is tightly locked at a tunable frequency difference away from a master laser (M), which is itself stabilised to an atomic transition in Cs. The frequency stability of the master laser is compared to a reference frequency provided by an optical frequency comb (C). The probe is spectrally filtered with an optical cavity (OC), and actively power-stabilised with an acousto-optic modulator (AOM).  The probe is then divided with high-precision into two beams using a Glan-Taylor (GT) polarizing prism followed by a Wollaston beamsplitter (W). One beam passes through a Cs cell  embedded inside a thermal and magnetic shield and is measured by photodiode A (PD A); the other beam is measured  directly by photodiode B (PD B). Temperature and thermal gradients are  measured independently with  calibrated standard  platinum resistance thermometers (SPRTs).
}

 \label{fig:setup}
\end{center}
\end{figure*}

At thermal equilibrium, the velocity distribution of atomic absorbers in a vapour cell is related to the temperature through the Boltzmann distribution. This simple and fundamental relationship forms an excellent foundation for  a type of primary thermometry known as  Doppler broadening thermometry (DBT) \cite{borde,borde2009}.  DBT differs from the current leading technique  for primary thermometry, which measures the temperature-dependent speed of sound in a noble gas contained in a well-characterized acoustic resonator.  Until recently, the best  determination of $\kB$ was made in 1988 with a total uncertainty of 1.7 parts-per-million (ppm) \cite{moldover}. Refinements to this technique over the last twenty years have reduced its uncertainty to the 1\,ppm level \cite{pitre2011,dePodesta2013},   improving the uncertainty in the CODATA-recommended value for $\kB$ from 1.8\,ppm (CODATA-02) to 0.91\,ppm (CODATA-10). Despite these superb measurements, it is important that different techniques are employed to measure $\kB$ to reduce the possibility of underestimating systematic uncertainties that may be inherent to any single technique \cite{RevModPhys.37.537}.

The DBT approach is one of the most theoretically transparent of  the new  approaches to primary thermometry.  The spectrum of a gas sample is measured with high precision, and a theoretical model is fitted to the measured data.  If the model includes all  of the relevant physics then one can accurately extract the contribution to spectral broadening  arising solely from the thermal distribution of atomic/molecular speeds. Ammonia probed by a frequency-stabilized CO$_2$ laser at 1.34\,$\mu$m was the first thermometric substance employed in a DBT experiment \cite{daussy2007direct}. Subsequently, an extended-cavity diode laser at 2\,$\mu$m was used to probe a ro-vibrational transition of CO$_2$ \cite{casa2008primary}. In these first experiments, the line-shape was assumed to be a Gaussian or a Voigt profile  (a Gaussian convolved with a Lorentzian). Since 2007, with the ambitious goal of approaching  1 ppm accuracy,  substantial  experimental and  theoretical improvements  have been made to DBT  using    ammonia \cite{Triki2012,Lemarchand2012}, oxygen \cite{cygan2010}, ethyne \cite{Hald2011}, and water \cite{Gianfrani2012}. However, one key challenge peculiar to  molecular absorbers is the need to account for complex collisional effects on the line shape~\cite{Lemarchand2012}.  We avoid  this  by using a dilute atomic vapour~\cite{Gar-Wing-Truong2010aa,truong2011} with a strong dipole transition for which a tractable, microscopic theory has been developed \cite{PhysRevA.86.012506,PhysRevA.81.033848}.

 The spectrometer  is  pictured in Fig.~\ref{fig:setup}.  The probe laser is spectrally and spatially filtered using a combination of an optical cavity of moderate finesse ($\mathcal{F}\approx305$) and single-mode fibre.  This reduces the spontaneous emission content of the probe beam from 1.6\% to below 0.01\%. It is then delivered into a vacuum chamber in which the vapour cell is mounted in a thermal and magnetic shield.  The temperature of the shield can be controlled to a few millikelvin and gradients are suppressed to the same level. The light is split into two output signals using a combination of a  Glan-Taylor polarising prism  and Wollaston beam-splitter. The ratio of powers in the output beams is stable to better than $10^{-6}$.  One output directly illuminates a photodiode to give the incident power, whilst the other passes  through the vapour cell and is then detected.  The ratio of these photodiode signals gives us  the transmission ratio, $\mathcal{T}$. The incident power is actively controlled while the frequency of the incident light is set to an absolute accuracy of 2\,kHz. The temperature of the vapour cell is monitored directly using a calibrated capsule-type standard platinum resistance thermometer (CSPRT), allowing us to quantify the performance of our Doppler thermometer. Further technical details about the apparatus are given in the supplementary information (SI).

\begin{figure}
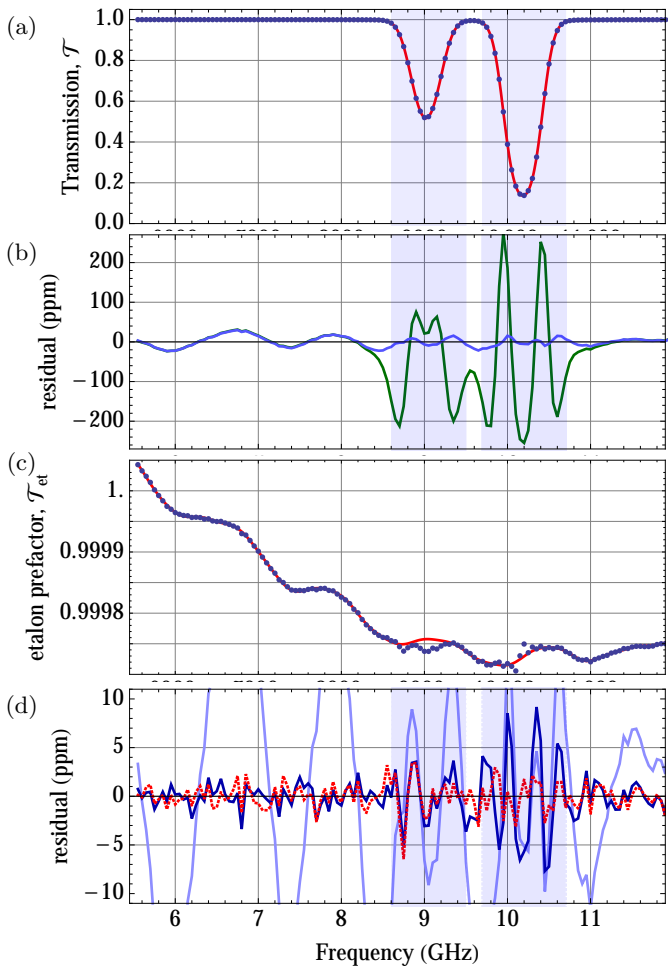

\begin{center}
\begin{lpic}[]{trans(8.6cm)}\lbl[t]{0,42;(a)}\end{lpic}
\begin{lpic}[]{res4bAndreGreenVoigtOnlyBlue1CorrNoetalons(8.6cm)}\lbl[t]{0,42;(b)}\end{lpic}
\begin{lpic}[]{etalon2327(8.6cm)}\lbl[t]{0,45;(c)}\end{lpic}
\begin{lpic}[]{res4dAndreLightBlue1Corr0EtalonDarkBlue1Corr6EtalonRed2Corr6Etalon(8.6cm)}\lbl[t]{0,55;(d)}\end{lpic}
\caption{(a) Measured transmission at 296 K for the highest intensity, $I/I_{\textrm{sat}}=0.0028$, averaged over 200 scans [points] and theory [solid]. (b) Residuals including Voigt only [dark-green] and first-order correction [light-blue].  (c) Etalon pre-factor, \eqn{prefactor}, with 6 etalons  [solid] and data divided by second-order power-corrected Voigt profiles [points],  assuming \mbox{$\LW=2.327$ MHz}. (d) Residuals after  including first-order correction with 0 etalons [light-blue] and 6 etalons [dark-blue], and second-order with 6 etalons [dotted-red]. 
} \label{fig:transmission}
\end{center}
\end{figure}

We measure the  transmission through the vapour cell as the probe laser frequency scans across  the Cs D1 transition (6S$_{1/2}$ -- 6P$_{1/2}$), shown in  Fig.~\ref{fig:transmission}(a).  The relative noise in the measurement of the atomic transmission is just 2 ppm in a 1\,s measurement (at the highest optical powers used). The two dips seen on  Fig.~\ref{fig:transmission}(a) come from the hyperfine splitting in the excited state and our excellent signal-to-noise ratio enables us to determine this splitting to be \mbox{$\HFS=1167.716(3)\textrm{ MHz}$} consistent with  previous measurements ($\HFS=$1167.680(30) MHz \cite{steck2003cdl,udem1999}), but 10 times more accurate. 

The absorption line-shape, $\mathcal{T}_\textrm{at}(f)$, of an atomic transition with a rest-frame transition frequency $f_{0}$, is given by a convolution of the natural (half) linewidth  $\LW= 1/(4 \pi\tau)$ due to the finite lifetime $\tau$ of the transition,   with a Gaussian distribution having a \mbox{$e^{-1}$-width} \mbox{$\DW= f_0\sqrt{2 \kB T/(m_\textrm{Cs} c^2)}$} due to  Doppler broadening   of  atoms of mass $m$ at temperature $T$. By fitting a line-shape to the measured transmission data, with $\DWfit$ as a fitting parameter, we extract the Doppler component,  from  which we infer the temperature.  Systematic errors in $\DWfit$ are quantified by  converting the independent PRT  temperature measurements into an expected Doppler width, $\DWprt$, {using the  CODATA values for $\kB$ and $m_\textrm{Cs}$}, which we compare with $\DWfit$.

  If the natural lineshape of the transition is a Lorentzian, then the resulting convolution is the well-known Voigt profile, which is commonly used to model dipole resonances.  However, our extremely low-noise transmission measurements reveal deviations from the Voigt profile.  Some of these deviations are technical in origin e.g.\ instrumental broadening due to the lineshape of the probing laser, residual spontaneous emission from the probe laser, unwanted optical etalons, and photodetector linearity; however, there is an important fundamental effect resulting from frequency-dependent optical pumping, which perturbs the atomic natural line-shape away from a Lorentzian.  All of these effects, whether technical or fundamental, cause systematic perturbations to the line shape, and must be taken into account    to model the lineshape accurately.  

\begin{figure}
\begin{center}
\includegraphics[width=\columnwidth]{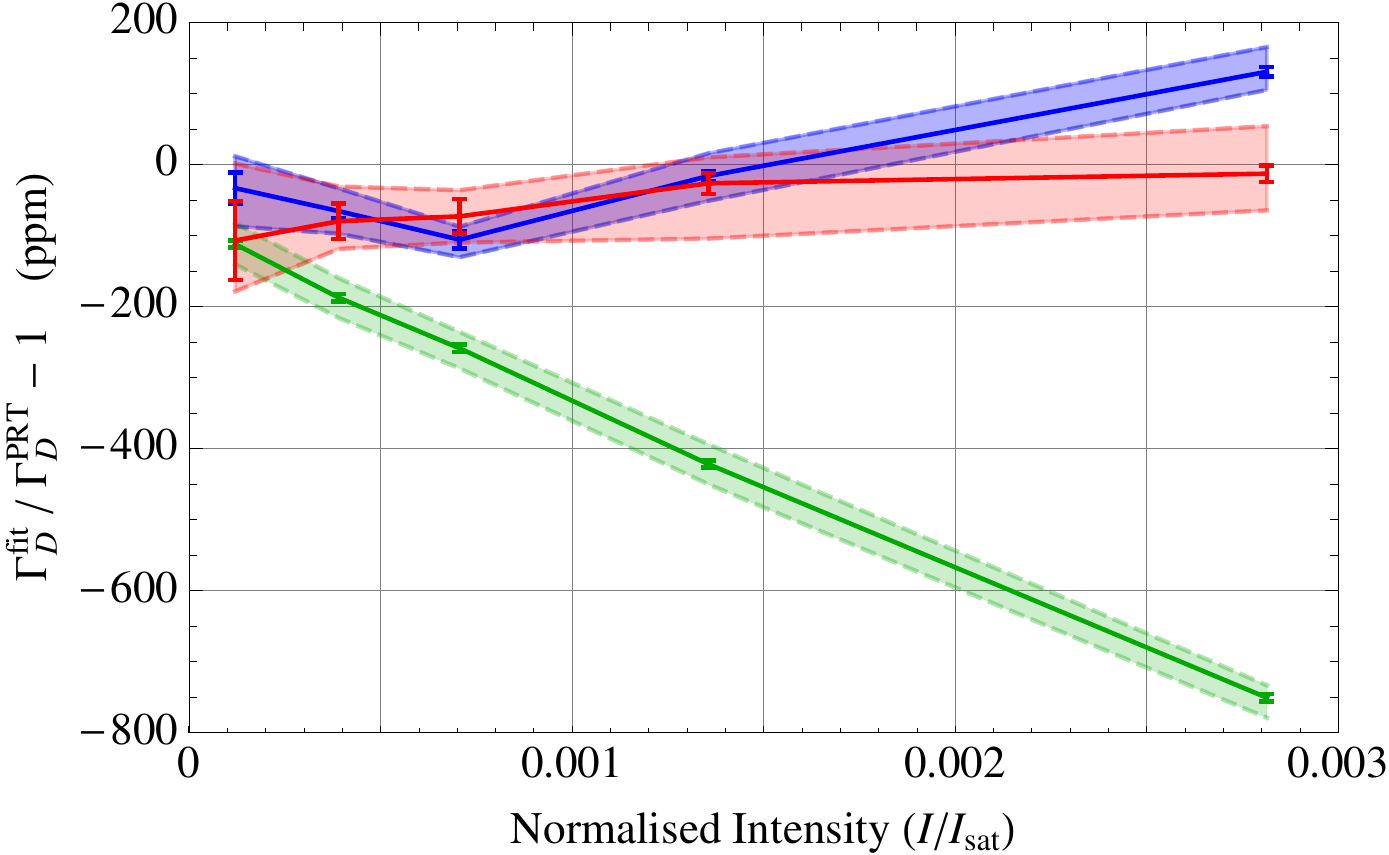}
\caption{Fractional deviation between the fitted Doppler width and that inferred from PRT temperature measurements at \mbox{296 K}.  Green points are fits using the power-independent Voigt-only profile.  Blue points are fits using the first-order intensity-dependent correction.  Red points are fits using the second-order intensity-dependent correction. Solid curves are fits using our central estimate of $\bar\LW=2.327$ MHz; dashed curves are for $\LW=2.320$ MHz (upper) and $\LW=2.334$ MHz (lower) representing $\bar\LW\pm\sigma_{\LW}$:  the shaded regions indicates the sensitivity of $\DWfit$ to $\LW$ in this range. Error bars represent  standard errors of fits to the average of  200 scans for fixed $\LW$.
} \label{fig:RoomTempDWPRT}
\end{center}
\end{figure}

To demonstrate these deviations, we fit a model consisting of two Voigt profiles separated by $\HFS$,  to the raw transmission data shown Fig.~\ref{fig:transmission}(a). The residuals of this Voigt-only fit have characteristic \textsf{M}-shaped features near resonance, with amplitude $\sim 200$ ppm at the highest probe powers, as shown in the green trace in Fig.~\ref{fig:transmission}(b). These features demonstrate the breakdown of the Voigt profile,  indicating additional, unaccounted physics,  which  causes a spurious linear dependence of $\DW^{\textrm{fit}}$ on probe intensity, as shown in the green curve on Fig.~\ref{fig:RoomTempDWPRT}. 
    In what follows, we describe how we sequentially  include additional physics in our transmission model to remove these, and other, systematic effects, leaving only    the shot-noise limit of our detection apparatus.
 
In earlier work we showed that the \textsf{M}-shaped features in Fig.~\ref{fig:transmission}(b) arise from atomic population dynamics induced by the probe laser, which are significant even at  exceedingly low  intensities \cite{PhysRevA.86.012506}.  We subsequently calculated corrections to the Voigt profile up to linear-order in the probe intensity \cite{PhysRevA.81.033848}, to account for optical pumping.  Including these first-order intensity-dependent correction in our model   suppresses the features substantially, as shown in the blue trace in Fig.~\ref{fig:transmission}(b).

 Far from the  resonances, oscillations are  evident in Fig.~\ref{fig:transmission}(b).  These arise from low-finesse etalons in the optics used to deliver the light to the atoms.  
The dominant etalon has an amplitude around 40\,ppm, corresponding to interference between surfaces with power reflectivities of $\lesssim10^{-9}$. Although obviously technical in nature, reducing the size of etalons beyond this already fantastically low level is an experimental challenge. Instead, we include  etalons in our transmission model, so that the total transmission is given by $\mathcal{T}=\mathcal{T}_\textrm{at}\mathcal{T}_{\textrm et}$, where  $\mathcal{T}_{\textrm et}$  includes $n$ etalons and a slowly-varying  quadratic background:
 \begin{equation}
\mathcal{T}_{\textrm et}(f)= \frac{\alpha +\beta f+\gamma f^2}{\prod_{j=1}^n \big(1+a_j \sin(2\pi f/f_j  +\phi_j)\big)}.\label{prefactor}
\end{equation}
Etalons are of particular concern for DBT since they  introduce systematic errors in $\DWfit$. We note that it is only because of our extreme transmission sensitivity that these features are revealed, and thus give the opportunity to   suppress this systematic error. 

We add etalons to the model until the residuals far from resonance are consistent with a  white-noise background. In practice, it is necessary to include up to $n= 6$ etalons for  high power scans, with the smallest  resolved etalons having amplitude \mbox{$\sim7$ ppm}.  The solid curve in Fig.\ \ref{fig:transmission}(c) shows $\mathcal{T}_{\textrm et}$ for the highest power data; the points show the raw transmission data divided by the fitted $\mathcal{T}_{\textrm at}$,    demonstrating that $\mathcal{T}_{\textrm et}$ is  uncorrelated with the atomic transmission profile shown in Fig.\ \ref{fig:transmission}(a).

\begin{table}%!]
\begin{center}
\renewcommand{\arraystretch}{0.7}
\begin{tabular}{l|l}
Source & {$u_r(k_B)$ (ppm)}	\\
\hline
\hline
Statistical (from \fig{fig:scanDW})  	
		&	\hphantom{1}5.8	
												\\
												\hline
Lorentz Width ($\LW^\textrm{fit}$)${}^a$		&	\hspace{8mm}65 
\\
\textit{Lorentz Width ($\LW^\textrm{ind}$)${}^b$} 		&	\hspace{12mm}\textit{190}	 \\
Laser Gaussian noise	&	16							\\
Optical pumping %@ 296\,K   	
& 	15						\\%15.4
 Etalons (misidentification)						&	15		\\%14.9
 Etalons (unresolved)						&	\hphantom{1}3					\\
Spontaneous Emission			&	\hphantom{1}3.6							\\
Temperature			&	\hphantom{1}1.9							\\
Temp.\ Gradient		&	\hphantom{1}1.2						\\
PD Linearity			&	\hphantom{1}1							\\
Zeeman Splitting		&	$<$1\hphantom{.2}							\\
\\[-3ex]
\hline
\hline
Total (fit $\LW$)    				&	\hspace{8mm}		71										\\
\textit{Total (independent $\LW$)} 			&	\hspace{12mm}\textit{191}	 											\\
\hline
\end{tabular}
\end{center}
\caption{Error budget for the determination of the Boltzmann constant at 296 K. 
Optical pumping shifts are reported at $I/I_{sat}= 3\times10^{-3}$ assuming second-order corrections to the Voigt profile. %
    ${}^a$Estimated from fits to scan data with free LW. %
${}^b$Estimated from published uncertainty in Cs lifetime, plus uncertainty in the independent estimate of the laser linewidth. $u_r$  denotes fractional uncertainty.
\label{budget}}
\end{table}

The dark blue trace in  Fig.~\ref{fig:transmission}(d)  includes  first-order power-dependent corrections, and $6$  etalons. Away from resonance, the residuals are then consistent with the  noise-floor of our  apparatus.    However, around the deepest resonance we  observe  \mbox{$\sim10$ ppm} features.    To eliminate these  we include second-order intensity-dependent corrections to the Voigt profile (see SI):
 \begin{eqnarray}
\mathcal{T}^{(i)}_{\textrm{at}}&=&e^{-p V_\nu(\Delta)}\Big(1+
q_1{(e^{-p V_\nu(\Delta)}-1) V_\nu^{(2)}(\Delta)}/{V_\nu(\Delta )}\nonumber\\
&&{}\hphantom{e^{-p V( }}+  q_2\big({(e^{-p V_\nu(\Delta)}-1) V_\nu^{(2)}(\Delta)}/{V_\nu(\Delta )}\big)^2\nonumber
   \\
  &&{}\hphantom{e^{-p V(}} + q_3{(e^{-2p V(\Delta ,\nu )}-1) V_\nu^{(3)}(\Delta)}/{V_\nu(\Delta)}\Big),
   \end{eqnarray}
   where $\nu=\DW/\LW$, $\Delta$ is the detuning from the resonance,  
  and the generalised Voigt profile is given by
$
V^{(n)}_\nu(\Delta)=\pi^{-3/2}\int_{-\infty}^{\infty} dx \,{e^{-{(\Delta-x)^2}/{\nu^2}}}{(1+{{x^2}})^{-n}},\label{eqn:V2}
$
and the conventional Voigt profile is $V_\nu(\Delta)=V^{(1)}_\nu(\Delta)$.   For a cell of length $L$ with linear absorption coefficient $\alpha$, the optical depth on resonance is $p=\alpha L$, \mbox{$q_1\propto I$} is the linear intensity-dependent coefficient \cite{PhysRevA.86.012506}, and \mbox{$q_{2,3}\propto I^2$} are quadratic intensity-dependent coefficients.  

The red dotted trace in Fig.~\ref{fig:transmission}(d) shows the residuals after including second-order intensity corrections.  The residuals are  suppressed to  $\sim2$ ppm, consistent with the detection  noise floor, giving  confidence that the transmission model accounts for  systematic effects.

\begin{figure}[t]
\begin{center}
\includegraphics[width=\columnwidth]{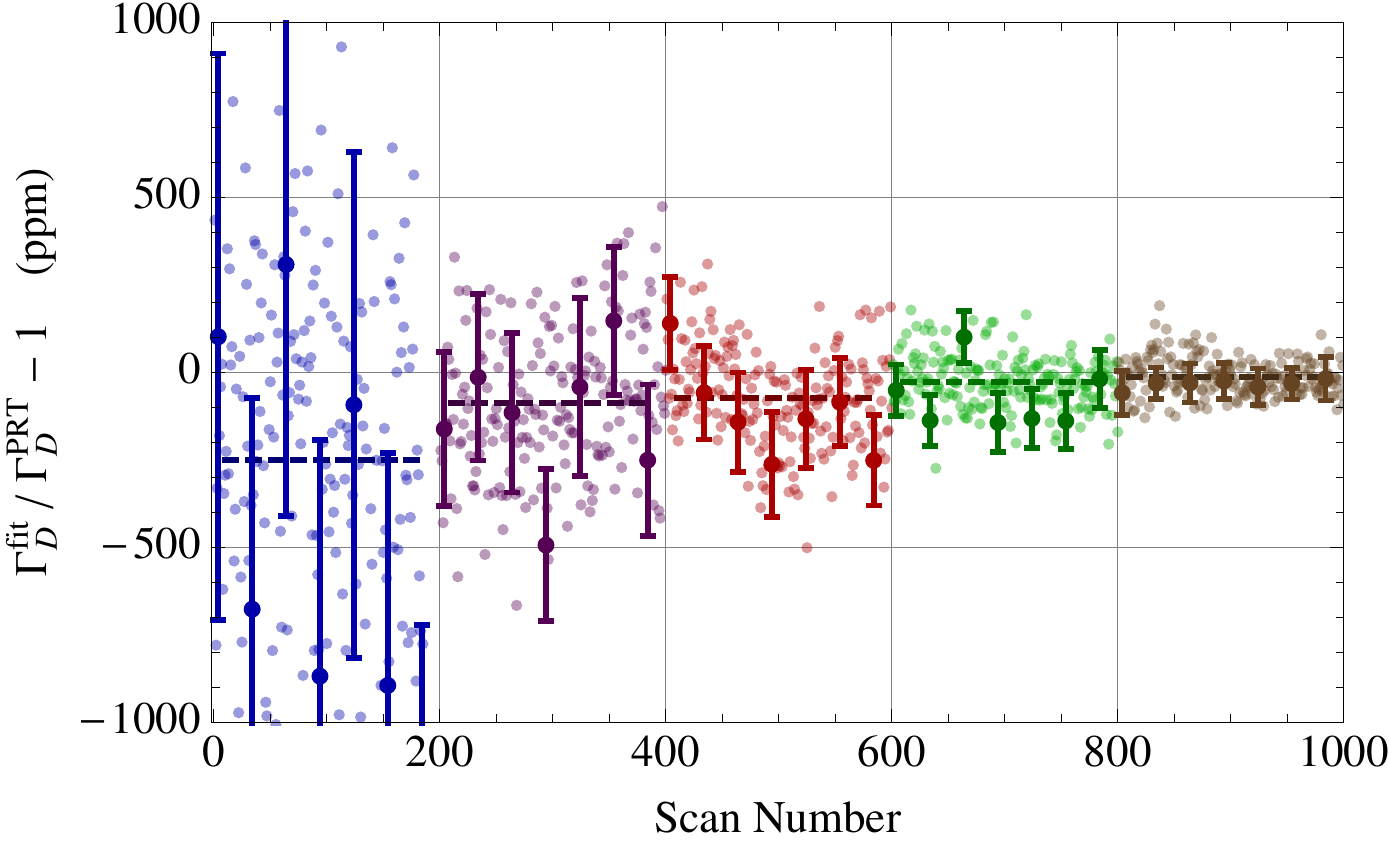}
\caption{Deviation between  $\DWfit$ (using second-order intensity correction) and $\DWprt$, for each scan at 296 K, with \mbox{$\LW=2.327$ MHz}.   Colours and vertical lines demark  different incident intensities (lowest at left, corresponding to normalised intensities in Fig.\ 1). Dashed lines are the mean of the corresponding set of scans, error bars are the estimated parameter error ($\pm1 \sigma$)  for each selected scan, and are consistent with sample standard deviation. Etalons are  included in each fit, with parameters fixed by fits to the average of all scans within a power level, i.e.\ etalon parameters are not free to vary within a particular power level.
} \label{fig:scanDW}
\end{center}
\end{figure}

 Figure~\ref{fig:RoomTempDWPRT} compares the fractional deviation between $\DWfit$ and $\DWprt$ as a function of incident intensity  for the three possible models for the atomic line shape: Voigt-only (green), first-order (blue) and second-order (red) intensity-dependent corrections. Naturally, the true $\DW$ should be intensity independent.  The simplest Voigt-only theory  clearly exhibits a linear intensity-dependence, leading to  $\sim 750$ ppm discrepancy even at intensities as low as \mbox{$I/\Isat=2.8\times10^{-3}$} where \mbox{$\Isat=2.5$\,mW/cm$^2$} is the saturation intensity for the transition \cite{steck2003cdl}. From \cite{PhysRevA.81.033848}, we expect the first-order correction to have a residual quadratic dependence on intensity, which is consistent with the blue curve on Fig.~\ref{fig:RoomTempDWPRT}.  Finally, the second-order correction (red curve) is seen to suppress  all intensity dependence to below the measurement precision.  The simultaneous removal of all systematic features from the residuals together with the elimination of intensity-dependence in the recovered $\DWfit$ gives a high degree of confidence that all relevant physics is properly included in the theoretical model.

\textbf{Thermometry}:
In this section we demonstrate the power of our spectroscopic technique by applying it to the problem of primary thermometry, and determine a  value for Boltzmann's constant  using atomic spectroscopy.  

The atomic vapour is brought to thermal equilibrium  at a chosen temperature by suspending it within a thermal isolator with an independent means for temperature measurement accurate to 0.55\,mK (1.9\,ppm). After systematically removing the effects of optical pumping and etalons, the  largest  source of uncertainty in $\DWfit$ comes from uncertainty  in  $\LW$.  We  obtain a value for  $\LW$ directly by fitting it to the data, which gives $\LW^\textrm{fit}=2.327(7)$ MHz.  This is consistent with, but also more precise than,  an independent estimate, \mbox{$\LW^\textrm{ind}=2.331(19)$ MHz}, given by the sum of the natural linewidth of the Cs $6P_{1/2}$ level, \mbox{$\LW^\textrm{at}=2.287(6)$ MHz} \cite{steck2003cdl}, and the laser linewidth, $\LW^\textrm{las}=0.044(18)$ MHz. From Fig.~\ref{fig:RoomTempDWPRT} we estimate that a 1\,kHz change in $\LW$ leads to $\sim5$ ppm change in $\DWfit$, so that the 6.5 kHz uncertainty in $\LW$ contributes 32.5 ppm uncertainty in $\DWfit$ [65 ppm in $u_r(\kB)$].  We note that if the probe laser linewidth were decreased to the kilohertz level, then our measurement would yield a state-of-the-art value for the excited state lifetime of Cs.

  We now briefly describe additional sources of uncertainty, which are quantified in Table \ref{budget} (see SI for a comprehensive description). 
Statistical error arising from the scatter in Fig.\,\ref{fig:scanDW} contributes 2.9 ppm to $u_r(\DWfit)$ [\mbox{5.8\,ppm} to $u_r(\kB)$].  This is also consistent with the estimated standard error for each point in Fig.\,\ref{fig:scanDW}, shown as error bars.  The probe laser has gaussian noise of width 0.88(39)\,MHz, contributing 8 ppm to $u_r(\DWfit)$ [16\,ppm to $u_r(\kB)$].  Residual effects of optical pumping after second-order power corrections contribute 7.5 ppm to $u_r(\DWfit)$, [15 ppm to $u_r(\kB)$].  Misidentification of etalon parameters contributes \mbox{$7.5$ ppm} to $u_r(\DWfit)$, [15 ppm to $u_r(\kB)$].  Possible unresolved etalons contribute 3 ppm to $u_r(\kB)$.  Residual spontaneous emission from the laser, after filtering by the optical cavity,  contributes \mbox{3.6\,ppm} to $u_r(\kB)$.

To determine a value for $\kB$, we  take a weighted mean of $\DWfit/\DWprt$ extracted from fits using all the second-order corrected data shown on Fig.~\ref{fig:RoomTempDWPRT} (red points).  We find \mbox{$\kB=1.380\,545(98)\times 10^{-23}$}\,J/K, where the 71 ppm  uncertainty is calculated in Table \ref{budget}. This  is consistent with the recommended CODATA  value of \mbox{$1.380\,648\,8(13)\times 10^{-23}$ J/K}.

In conclusion, we have developed an atomic absorption spectrometer that operates with a transmission measurement precision of just 2\,ppm. This has revealed several new phenomena.  In combination with a theory that correctly treats the interaction between light and an effusive vapour we are able to explain all of the observed effects to a level of precision never before demonstrated. The power of our technique is  demonstrated by measuring the line-shape parameters of a gas at thermal equilibrium.  Our reproducibility is exactly consistent with the quantum-limits of measurement \cite{PhysRevA.82.011611} which gives us confidence that we have captured all of the relevant physics.  

With our unprecentented sensitivity, we have measured the transmission spectrum of Cs at 895\,nm at the shot-noise limit. 
From these measurements, we derived a value for Boltzmann's constant with an uncertainty of 71 ppm, which is consistent with the recommended CODATA value, and the hyperfine frequency splitting of the $6^2P_{1/2}$ level with an uncertainty ten times better than ever previously  reported: 
\begin{eqnarray}
\kB &=& 1.380\,545(98)\times 10^{-23}\textrm{ J/K} \nonumber\\
\HFS &=& 1167.716(3)\textrm{ MHz.}\nonumber
\end{eqnarray}
Our current performance permits a measurement of the Boltzmann constant with a precision of 6\,ppm after only a few hours of data acquisition.  Our total uncertainty is  dominated by the imprecision in the literature value of the excited state lifetime of the Cs D1 transition (known to 0.26\% \cite{steck2003cdl}). A future experiment designed to further reduce the amplitude of etalons, and using   a more broadly tuneable ($\sim 30$\,GHz) and  higher power probe laser with sub-kHz linewidth  would eliminate probe laser effects,  allow better estimation of  etalon contamination, and lower the contribution of optical pumping.  This     opens the door  to the best measurement of the Cs D1 transition  lifetime, and a ppm-level measurement of $\kB$   thus making an important contribution to efforts to redefine the kelvin. 

The authors wish to thank the NIST Precision Measurement Grants Program, the Australian Research Council through the CE110001013, FT0991631 and DP1094500 grants for funding this work. The authors also wish to acknowledge the South Australian Government who have provided generous financial support through the Premier's Science and Research Fund. We thank Michael Moldover and Greg Strouse of NIST, and Mark Ballico from NMI Australia, for their contributions to the SPRT thermometry.

%\bibliography{bib}

\end{document}

\title{
Supplemental Information: Quantum-Limited Spectroscopy
}

\date{\today}

\pacs{
}

\maketitle

\section{Apparatus}
\begin{figure}
\begin{center}
\includegraphics[width=\columnwidth]{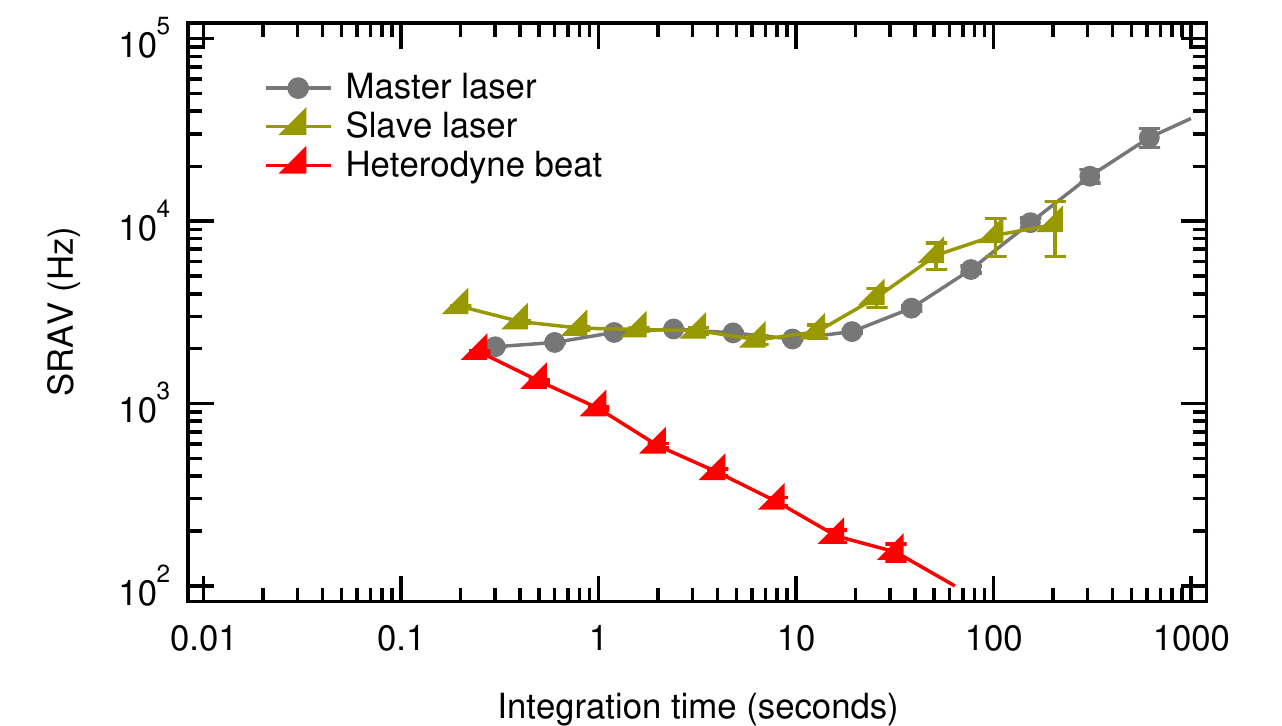}
\caption{Square-root-Allan-variance (SRAV) of the absolute frequency stabilities of the master laser (grey circles), slave laser (olive triangles) and the heterodyne beat (red triangles). 
} \label{fig:srav}
\end{center}
\end{figure}

\textbf{Light Source}: %
A dual-beam linear absorption spectrometer operating at 895\,nm was built using a pair of extended cavity diode lasers (ECDLs). The first of these, the master laser, is locked to the $6S_{1/2}$ $F=4$ to $6P_{1/2}$ $F=3$ transition of the \CsD{1} line in Cs at 894.6054\,nm. This provided an optical reference frequency with a square-root-Allan-variance (SRAV) of $\sim$2\,kHz for timescales between 1\,s and 30\,s, as shown by the grey circles on Fig.\,\ref{fig:srav}. These are the relevant time-scales for the spectroscopy, reflecting the time to take a single scan.  The other laser is  frequency locked to the master at a user-selectable frequency offset by stabilising the heterodyne beat frequency against a tunable radio frequency (RF) reference oscillator. This frequency lock contributed no additional frequency instability to this slave laser (red triangles). The resulting slave laser frequency stability is shown by the olive trace on Fig.\,\ref{fig:srav}. The  RF reference oscillator was step tuned under computer control to ensure that the  transmission data was synchronous with the selected frequency.  The wavelength of the slave laser was centered on the twin absorption peaks at 894.5793\,nm, corresponding to the $6S_{1/2}$ $F=3$ to $6P_{1/2}$ $F=3$ and $F=4$ transitions. The offset frequency was tuned in increments of 50\,MHz over a 6500\,MHz span that captured the atomic absorption features.

The slave laser output was coupled into a \FP resonator whose length was actively controlled to keep it resonant with the slave laser light. The output of the cavity was coupled into an acousto-optic modulator (AOM) which was used to actively stabilise the power in the deflected beam.  The output of the AOM was coupled into a single-mode fibre for spatial filtering and for delivery into the vacuum chamber.  This fibre contained an auxiliary detector on a fibre coupler which provided the signal to stabilise the input power into the chamber. The rest of the  light in the fibre was sent to a Cs sample cell embedded in a thermal and magnetic shield. 

\textbf{Spontaneous Emission}: %
In a separate experiment the slave laser light was tuned to the centre of the atomic resonance and passed through a heated Cs cell for which the optical depth was  30 i.e. $\Tatoms\approx e^{-30}\approx10^{-12}$.  A measurement of the residual transmission demonstrated that the laser output  contained  1.6\% broadband emission at its operating point. 

The laser light was filtered using the fundamental mode of a scannable \FP cavity ($\mathcal{F}=$305 and free-spectral range 17.6\,GHz).  This was necessary because broadband emissions contaminate the  measurement by creating a transmission offset associated with the non-resonant photons. Low level feedthrough of light from higher-order spatial modes of the optical cavity ($\sim$ 20\% of the main mode) was heavily attenuated by coupling the beam into a single-mode optical fibre prior to introduction into the evacuated chamber. 

  The combined cavity and single-mode filtering provides a reduction in the spontaneous emission by a factor $2\mathcal{F}/\pi\approx195$. We numerically modeled the effect of the residual spontaneous emission (0.008\%) on the measured Cs spectrum and found that the perturbation to the fitted atomic Doppler width $\nu_D$  was only 1.8\,ppm (equivalent to 3.6\,ppm in $u_r(\kB)$). This systematic error can easily be reduced to below the 1\,ppm level by a cavity finesse of over 1000.

\textbf{Instrumental linewidths}: %
The probe laser spectral lineshape can be expressed to a high degree of accuracy as a Voigt function \cite{didomenico2010} with a Lorentzian half-width-at-half maximum of $\LLW$ and Gaussian $e^{-1}$-halfwidth of $\LGW$. When this instrumental function is convolved with the atomic absorption profile (which, to first order, is also a Voigt function with parameters $\LW$ and $\DW$), the  apparently observed atomic lineshape is also a Voigt function with Lorentzian and Gaussian components given by $\LW=\LW^{\textrm{at}}+\LLW$ and $\Gamma_D=\sqrt{(\DW^{\textrm{at}})^2+(\LGW)^2}$, respectively.  

The  power spectral density (PSD), $S_\nu(f)$, of the probe laser  frequency noise was directly measured by using the frequency-dependent absorption on the side of an atomic transition.  Two regions of the PSD can then be identified \cite{didomenico2010}; at higher Fourier frequencies we see a dominant white frequency process, while at lower frequencies we observe steeper noise which in our case has a dominantly flicker ($1/f$) character. White noise components (with amplitude $S_0$)  produce a Lorentzian lineshape with HWHM given by $\pi S_0/2$. The integrated probe laser noise at lower frequencies give rise to a Gaussian lineshape which has an estimated   e-fold width of $\LGW=0.88(39)$\,MHz when integrated over the 40\,ms observation time for each frequency point in the scan.

\textbf{Optical power control}: %
The spectrally purified light is passed through an acousto-optic modulator (AOM) and the deflected beam is coupled into a non-polarising fiber splitter, which splits the power in the ratio 9:1. The lower power beam is measured on a photodiode and a feedback loop adjusts the AOM input to maintain a constant optical power in the fibre. The higher power output is out-coupled and used for spectroscopy. The polarisation of this beam is fixed by passing it through a Glan-Taylor prism, and then a Wollaston beamsplitter is used to divide it into a reference and sample beam. The sample beam passes through the Cs vapour and common-mode amplitude variations in this signal are removed by dividing it by the reference beam signal. 

\textbf{Conventional thermometry}: %
The Brewster-angled Cs sample cell is embedded in a cylinder of copper inside a multilayered thermostat, which is itself inside an evacuated chamber. The temperature of the thermostat can tuned using a thermo-electric cooler attached to the outer-most layer of shielding. The temperature at the cell was monitored using a capsule-type standard platinum resistance thermometer (CSPRT) that was calibrated to the ITS-90 temperature scale with a temperature precision of better than 0.5\,ppm in the range between 273\,K and 300\,K. However, our resistance meter (HP Model 3458A) limited our temperature measurement uncertainty to 1.2\,ppm. Measurements using a second calibrated CSPRT located in the copper cylinder at a distance $\sim50$\,mm away from the primary CSPRT revealed no temperature gradients above the 1.2\,ppm level. 

%ITS-90 vs thermodynamic T correction.
The ITS-90 temperature scale, $T_{90}$, itself is known to deviate from the true thermodynamic temperature, $T$ by $T-T_{90}=3.2\pm$0.4\,mK at $T_{90}=296$\,K\cite{fischer2011}. When the uncertainty in this correction is included, the total uncertainty due to conventional thermometry is 1.9\,ppm.

\textbf{Magnetic Shielding}: %
The magnetic environment was nulled around the Cs cell using a dual-layer shield made from high permeability metal. Numerical modeling and experimental characterisation of this shield demonstrated that the apparent broadening of the atomic absorption line due to Zeeman splitting of the otherwise degenerate levels contributed less than 1\,ppm uncertainty in $k_B$.

\begin{figure}
\begin{center}
\includegraphics[width=\columnwidth]{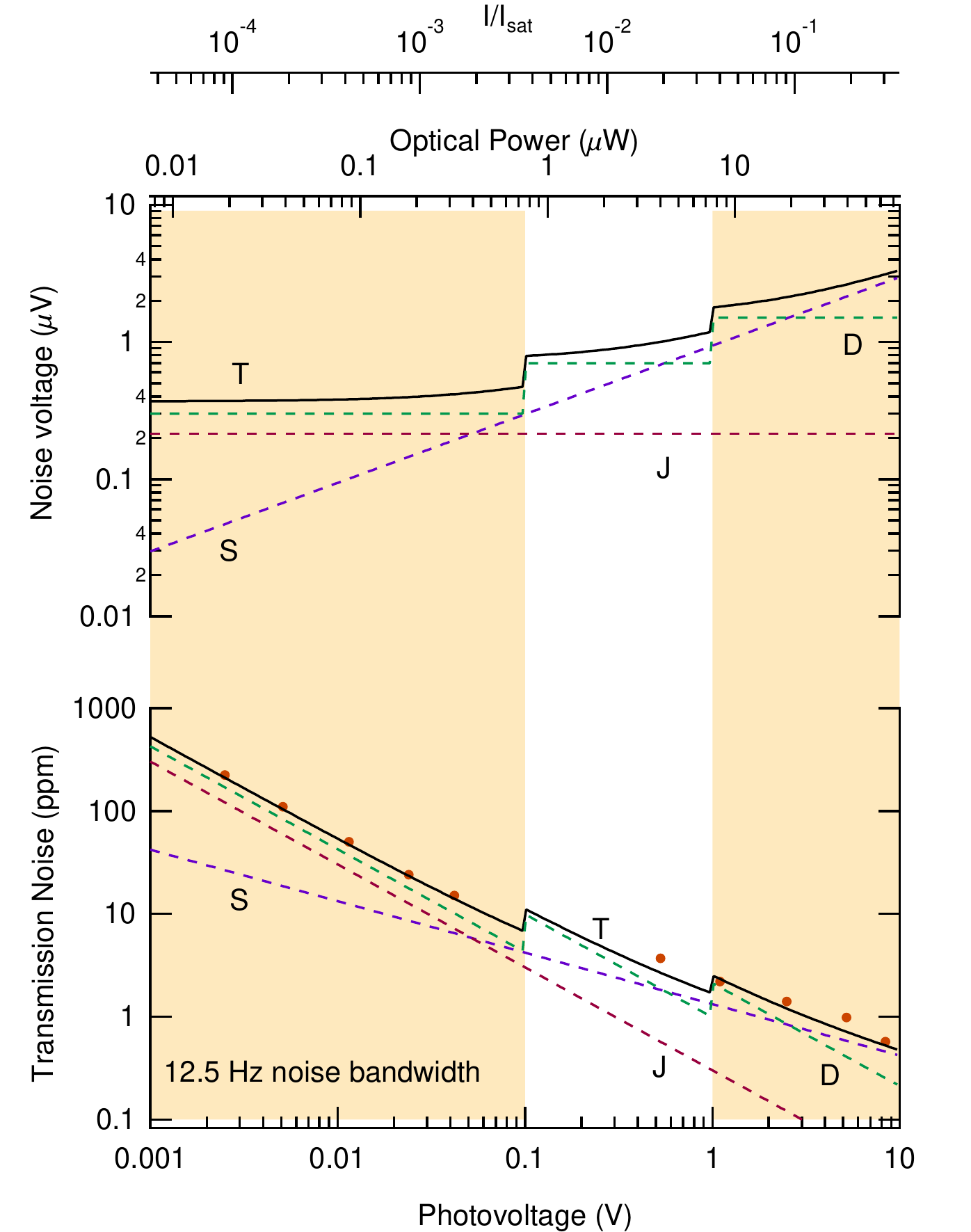}
\caption{
Top: Calculated shot noise (S) and Johnson noise (J) voltages as a function of the photosignal. The DMM (D) input noise was empirically determined and found to depend on the measurement range setting (indicated by the shaded regions).  Bottom: Measured noise in the atomic transmission channel (circles) compared to the calculated (with no free parameters) contributions from the DMM, Johnson noise, shot noise and total (T) are also shown for comparison.
} \label{fig:noise}
\end{center}
\end{figure}

\textbf{Optical detection}: %
Reverse-biased silicon photodiodes were used to measure the optical power in each of the reference and sample beams. The photosignals were measured by digital multimeters (DMMs) recording the voltage generated by the photocurrent passing through a load resistor. The noise contributions to each measurement of the atomic transmission from shot noise, Johnson noise and the DMMs are shown in Fig\,\ref{fig:noise} as a function of input signal level to each photodiode (the stair-case structure revealed in the figure arises from a change in the DMM range). It is clear that the transmission measurement are shot-noise limited in several bands of powers that lie just near the maximum of each range. At the highest powers used in the experiments ($\sim$0.6$\mu$W), shot noise  contributed more than $50\%$ of the total measurement noise. 

 The linearity of the photodetection chain was measured in two ways. In the first approach, the ratio of the transmitted and incident detector signals were measured over the full range of common-mode input power variation. This provided a reliable calibration curve for which the deviations from linearity at high powers could then be corrected. For low powers, this method did not provide sufficient signal-to-noise for calibration. Instead, we monitored the ratio of absorption coefficients for the two optical transitions. The second approach relies on the fact that the Clebsch-Gordon coefficients for these transitions require that the linear (i.e.\ low-power)  ratio of absorption coefficients should be 1:3 \cite{steck2003cdl}.  Forming this ratio, and enforcing that the extrapolation to zero power agrees with the 1:3 condition, we have a guaranteed, \textit{in situ} and independent way to determine any nonlinearity in the detector response at low powers. In  this way, we were able to resolve and remove the nonlinearities in the photodetection scheme. The technique allowed us to measure offsets corresponding to \mbox{$<$8\,pW} allowing us to ensure linearity  over a dynamic range of \mbox{$10^5:1$}. A more complete description of this method is found in Ref.\,\cite{truong2014linearity}.

\section{Error Budget}
Here we give a detailed description and derivation of the uncertainties reported  in Table I of the main text. $u_r(X)$  denotes the fractional uncertainty in $X$.

\begin{figure}
\begin{center}
\includegraphics[width=\columnwidth]{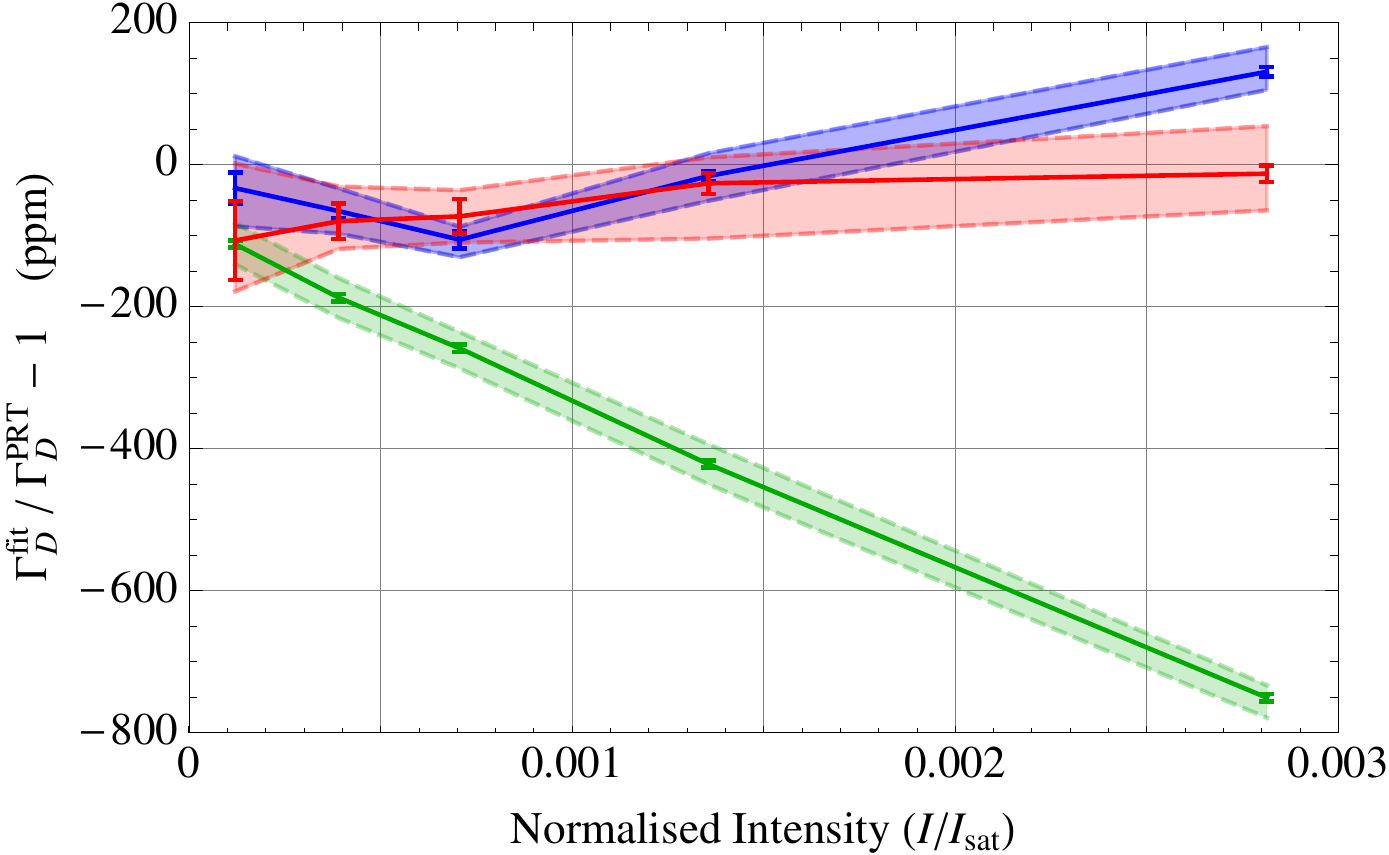}
\caption{Fractional deviation between the fitted Doppler width and that inferred from PRT temperature measurements at \mbox{296 K}.  Green points are fits using the power-independent Voigt-only profile.  Blue points are fits using the first-order intensity-dependent correction.  Red points are fits using the second-order intensity-dependent correction. Solid curves are fits using our central estimate of $\bar\LW=2.327$ MHz; dashed curves are for $\LW=2.320$ MHz (upper) and $\LW=2.334$ MHz (lower) representing $\bar\LW\pm\sigma_{\LW}$: the shaded regions indicates the sensitivity of $\DWfit$ to $\LW$ in this range. Error bars represent  standard errors of fits to the average of  200 scans for fixed $\LW$.
} \label{fig:RoomTempDWPRT}
\end{center}
\end{figure}

\begin{figure}[t]
\begin{center}
\includegraphics[width=\columnwidth]{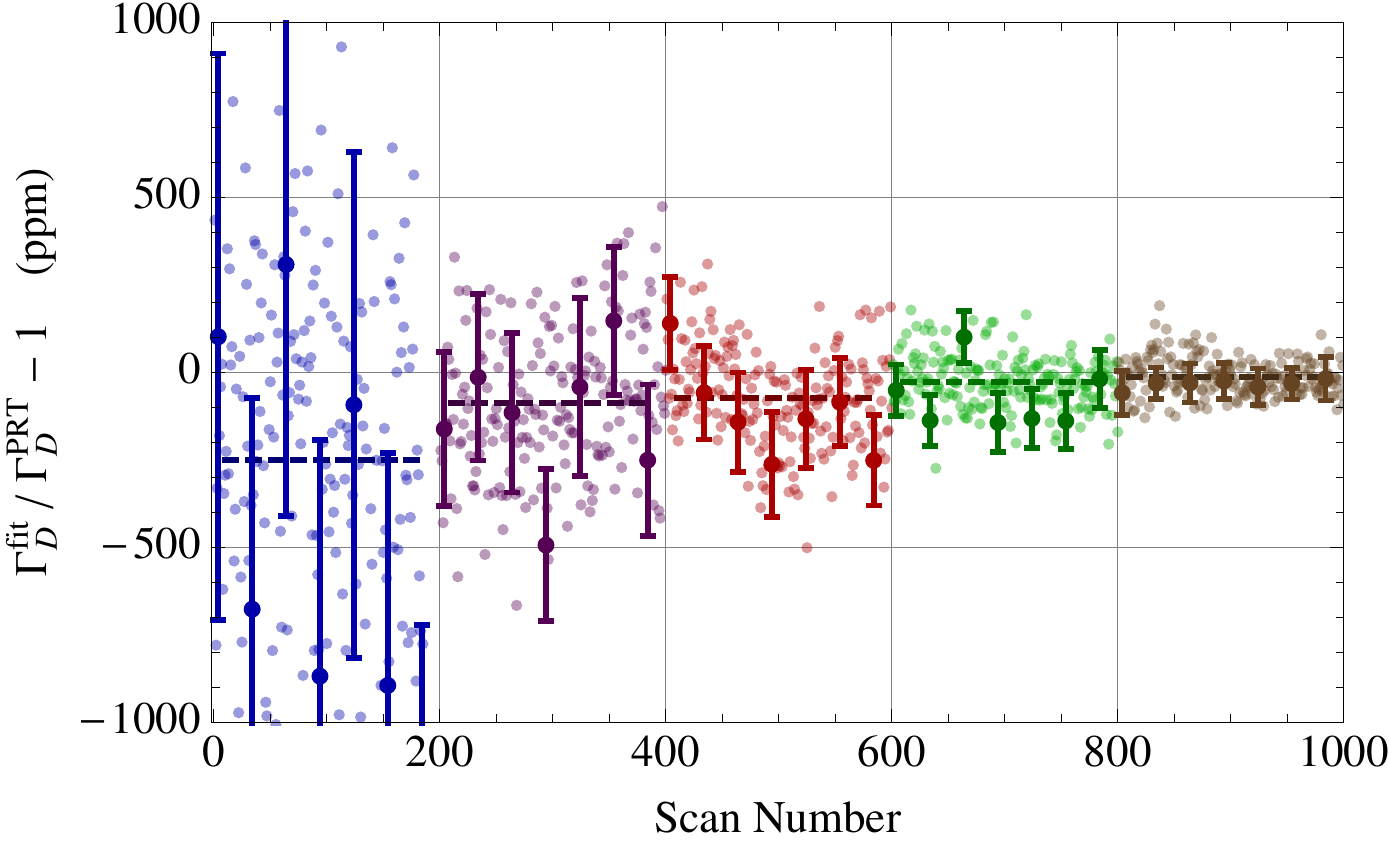}
\caption{Deviation between  $\DWfit$ (using second-order Voigt correction) and $\DWprt$, for each scan at 296 K, with \mbox{$\LW=2.327$ MHz}.   Colours and vertical lines demark  different incident intensities (lowest at left, corresponding to normalised intensities in Fig.\ 1). Dashed lines are the mean of the corresponding set of scans, error bars are the estimated parameter error ($\pm1 \sigma$)  for each selected scan, and are consistent with sample standard deviation. 
} \label{fig:scanDW}
\end{center}
\end{figure}

\textbf{Statistical}:  For each power level, the statistical contribution is computed by fixing the etalon parameters (to values obtained from fits to coaveraged scans, see below), and extracting $\DWfit$ from each scan independently. These are shown in Fig.\,\ref{fig:scanDW}, and we find a statistical error of 2.9\,ppm in $\DWfit$, [which represents 5.8 ppm in $u_r(\kB)$].

\textbf{Lorentz Width}: Our probe laser has a Lorentzian linewidth, which adds to the atomic linewidth, $\LW=\LW^\textrm{at}+\LW^\textrm{las}$.  To determine $\LW$, there are two alternatives: either to treat  $\LW$  as a free fitting parameter (together with  $\DW$),  or to fix $\LW^\textrm{at}$ to the best reported literature value and independently estimate $\LW^\textrm{las}$. 
  
   In the former route we fit to the average of 200 independent scans for each probe power level we  retrieve a value of $\LW^\textrm{fit}=2.327(7)$\,MHz.  
    In  the latter route, we take  $\LW^\textrm{at}=2.287(6)$ MHz  for the natural linewidth of the Cs $6P_{1/2}$ level \cite{steck2003cdl}, and an independent estimate of the laser noise  $\LW^\textrm{las}=0.044(18)$ MHz (see above), which totals $\LW^\textrm{ind}=2.331(19)$ MHz. These two estimates for $\LW$ are consistent within their measurement uncertainty, and so we take the value with smaller uncertainty \mbox{$\LW=2.327(7)$\,MHz}.  In particular, the solid curves in Fig.~\ref{fig:RoomTempDWPRT} assume the central value $\LW=2.327$\,MHz, and the dashed curves correspond to the upper and lower bounds of the uncerainty range, $\LW=2.320$\,MHz and $2.334$\,MHz.

We note that the relative precision in our measurement of $\LW$ is within a factor of two of the precision of the best measurements of the Cs D1 transition lifetime, $\tau_\textrm{Cs}$ \cite{PhysRevA.50.2174,PhysRevLett.91.153001}, and matches the precision of the  best synthesised  value of $\tau_\textrm{Cs}$ \cite{steck2003cdl} formed by combining these measurements.

We also see from Fig.~\ref{fig:RoomTempDWPRT} that a 1\,kHz variation in $\LW$ leads to $\sim5$ ppm change in $\DWfit$, {which is consistent with a 1:1 tradeoff between increases in $\LW$ and decreases in $\DWfit$, when $\DW\approx215$ MHz}.  This implies that our \mbox{$\pm6.5$ kHz} uncertainty in $\LW^\textrm{fit}$ contributes $32.5$ ppm to $u_r(\DWfit)$  (equivalent to 65 ppm to $u_r(\kB)$).  The $\pm19$kHz uncertainty in $\LW^\textrm{ind}$ represents 190 ppm in $u_r(\kB)$.

\textbf{Laser Gaussian Noise}: 
The probe-laser spectrum  has a Gaussian  contribution \cite{didomenico2010}. This  was measured  and found to have a   width of 0.88(39)\,MHz. The Gaussian component sums in quadrature with $\DW$, and so contributes 8 ppm to $u_r(\DWfit)$ [16\,ppm to $u_r(\kB)$]. 

\textbf{Optical Pumping}: 
The second-order power-dependent corrections suppress the  fit residuals in (shown in  Fig.~2(d) of the main text) to the measurement noise floor, so there is no statistically significant signature of optical pumping.  Nevertheless, any residual intensity dependence in the second-order intensity-corrected $\DWfit$ is expected to scale as $(I/I_\textrm{sat})^3$.  We therefore estimate the error arising from uncorrected third-order optical pumping by fitting a cubic through the highest intensity red data points in \fig{fig:RoomTempDWPRT}.  This gives a maximum deviation of 15 ppm in $u_r(\DWfit)$ at the \emph{maximum} probe intensity, and smaller contributions at lower intensities. Forming the weighted average over all intensities in \fig{fig:RoomTempDWPRT} contributes 7.5 ppm to $u_r(\DWfit)$, [15 ppm in $u_r(\kB)$]. 

\textbf{Etalons}: %
Etalons enter into the uncertainty budget in two ways. Firstly, etalon parameters may be misidentified during fitting, and so contaminate $\DWfit$.  Secondly  unresolved etalons masked by measurement noise may introduce residual systematic shifts to $\DW$. 

In Fig.\,\ref{fig:RoomTempDWPRT} we fit $\DWfit$ with free etalon parameters.  The weighted average of the standard errors is 8\,ppm in $u_r(\DWfit)$, which is a convolution of the statistical noise discussed above and  uncertainties arising from possible misidentification of the etalon parameters during fits.  When the statistical errors in $\DWfit$ are deconvolved from the standard errors in Fig.\,\ref{fig:RoomTempDWPRT}, we find the systematic error due to free etalon parameters in the fitting routine to be \mbox{$(8^2-2.9^2)^{1/2}= 7.5$ ppm} in $u_r(\DWfit)$, [15 ppm in $u_r(\kB)$].

We bound the error arising from unresolved etalons in two ways: by synthesising data with false etalons, and by computing the fractional shift in $\DW$ after the addition of the $j^\textrm{th}$ etalon in the fit function. Both approaches give worst-case shifts in $\DWfit$ that are comparable to the amplitude of the etalon.  The largest unresolved etalons are smaller than the measurement noise floor, so these cannot contribute more than 3 ppm to $u_r(\kB)$. 

\section{Lineshape Corrections at Second Order in Intensity}

\newcommand{\erfc}{\text{erfc} }
\newcommand{\re}{\text{Re} }

Here we derive Eq.\ 2 of the main text, which gives the second-order intensity-dependent correction to the Voigt profile.  We follow closely the derivation in \cite{PhysRevA.81.033848}.

We begin by computing the spectral dependence on the atomic populations for a three level atom, consisting of two ground states: one optically active, $\ket{1}$, and the other optically inactive, $\ket{2}$, and an excited state $\ket{3}$.  Transitions between states $\ket{1}$ and $\ket{3}$ are optically driven, and state $\ket{3}$ can relax to either of the ground states.   The population rate equations are \cite{PhysRevA.81.033848}
\begin{equation}
\dot{ \vec{P}}=M\vec{P},\hspace{0.25cm} M=\left(\begin{array}{ccc}\frac{\Omega^2f(t)/2}{1+\Delta ^2} & 0 & 2\beta+\frac{\Omega^2f(t)/2}{1+\Delta ^2} \\
0 & 0 &2(1-\beta)\\
-\frac{\Omega^2f(t)/2}{1+\Delta ^2}& 0 & -2-\frac{\Omega^2f(t)/2}{1+\Delta ^2}\end{array}\right),\label{eqn:rate}
\end{equation}
where $\vec{P}=\{P_1,P_2,P_3\}$, $\Delta$ is the detuning between the laser frequency and the atomic transition, $\beta$ is the branching ratio from state $\ket{3}$ to $\ket{1}$, $\Omega$ is the peak atomic Rabi frequency (proportional to the electric field amplitude) and $\sqrt{f(t)}$ represents the temporal profile of the Rabi frequency experienced by the atom as it moves through regions of different intensity.  We write the Rabi frequency in this form since in what follows we will expand the atomic response in powers of ${\Omega^2}/({1+\Delta ^2})$.  For simplicity of presentation, we will assume that the equilibrium atomic level populations are $\vec{P}_{\mathrm{eq}}=\{1/2,1/2,0\}$, i.e.\ we ignore degeneracies in the ground state manifold.  Accounting for degeneracies is a simple modification to the relative fractions for $P_1$ and $P_2$ at equilibrium.

\begin{figure}[t]
\begin{center}
\includegraphics[width=\columnwidth]{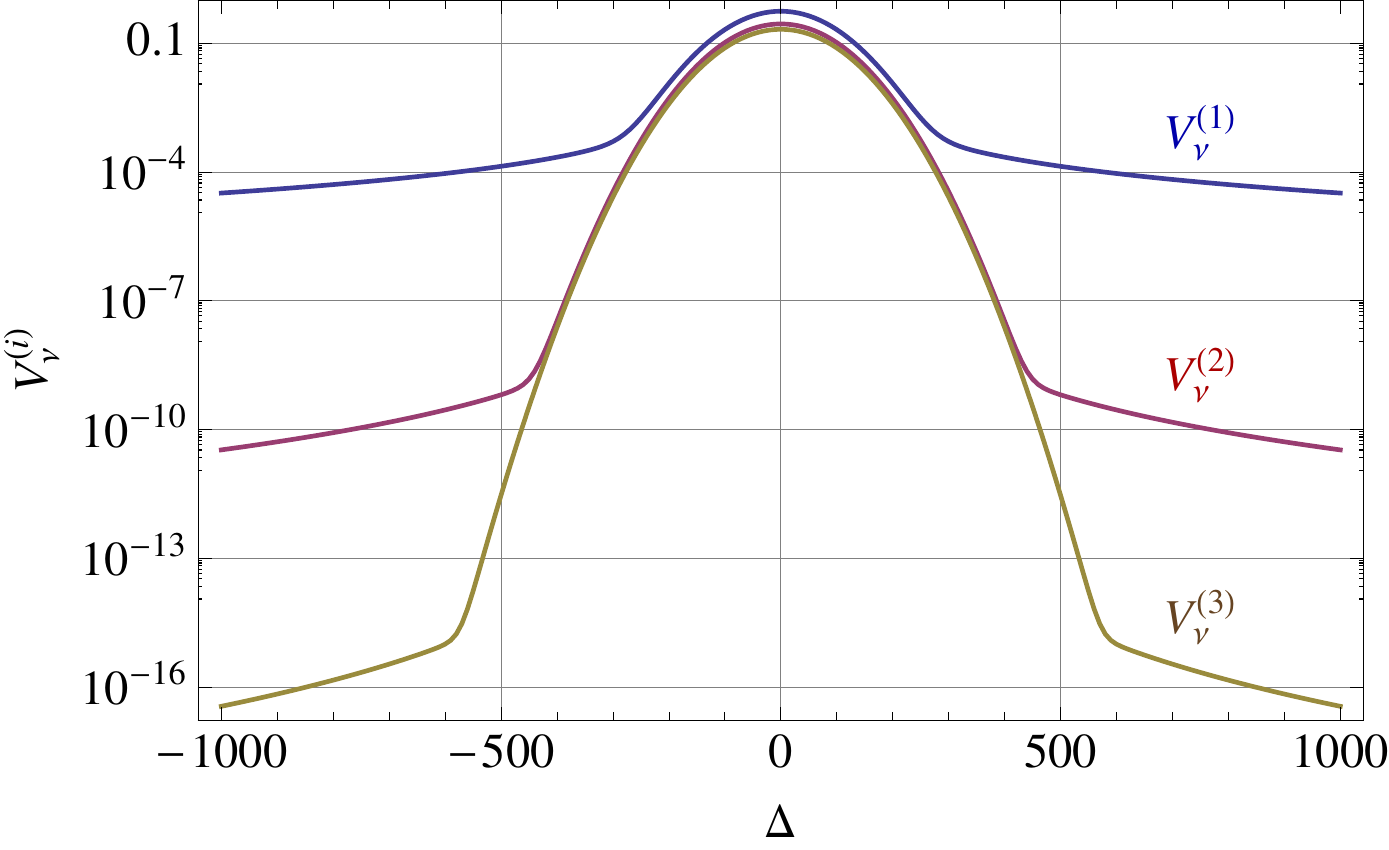}
\caption{Generalised Voigt profiles versus detuning, $\nu=100$.
} \label{fig:Voigt}
\end{center}
\end{figure}

The light is absorbed as it propagates through the atomic medium.  While \eqn{eqn:rate} is defined with respect to the rest-frame of a particular atom,  the axial evolution of the field in the lab-frame is governed by \cite{PhysRevA.81.033848}
\begin{equation}
\frac{\partial\,\ln\Omega^2}{\partial z}
=2\,\pi^{3/2} K \int_{-\infty}^{\infty} d\Delta_{v_z} \frac{e^{-{(\Delta-\Delta_{v_z})^2}/{\nu^2}}}{\pi^{3/2}(1+{{\Delta_{v_z}^2}})}P(\Delta_{v_z}),\label{eqn:integral}
\end{equation}
where $P=P_3-P_1  $,
  \mbox{$ K ={\rho_0\mu^2}/({\sqrt{\pi}\varepsilon_0\hbar v_0})$}, $\mu$ is the transition dipole moment and $v_0=\sqrt{2k_BT/m}$ is the mean thermal atomic velocity for an atom of mass $m$.  The notation $\Delta_{v_z}$ indicates that a given atom is subject to a detuning which depends on its axial velocity via the Doppler shift.  If the field intensity is weak, so that on resonance populations are negligibly perturbed from thermal equilibrium, then $P=-1/2$, and the integral yields $\Omega^2(z)=\Omega^2_0 e^{-\pi^{-3/2} K  z V_\nu(\Delta)}$.  Since $\mathcal{T}(z)=\Omega^2(z)/\Omega^2_0$, we recover $\mathcal{T}=e^{-\pi^{-3/2} K  z V_\nu(\Delta)}$, which is the conventional Beer's law result for a Voigt atomic lineshape.

For larger $\Omega^2$, perturbations to $P$ become significant, and we need to compute corrections to $P$ arising from the dynamics described by \eqn{eqn:rate}.    Since we are concerned with small but significant pump induced perturbations to the thermal population, we do a perturbative analysis of \eqn{eqn:rate} around the limiting case where $\Omega=0$.

By inspection of \eqn{eqn:rate} we see that $\Omega^2$ only ever appears as part of the ratio ${\Omega^2}/({1+\Delta ^2})$, so it must appear in the same ratio in $P$, which expand in powers of $\Omega^2$ \cite{PhysRevA.81.033848}:
\begin{equation}
P(\Delta)=a_0+a_1 \frac{\Omega^2}{1+\Delta^2}+a_2\left( \frac{\Omega^2}{1+\Delta^2}\right)^2+O(\Omega^6),\label{eqn:sol}
\end{equation}
 where  $a_0=-1/2$ and other $a_j$ are expansion coefficients, which contain all the complexity of the ballistic atomic motion across the beam, including averages over  transverse velocity classes and trajectories across the beam.  Here $P$ is defined in the lab-frame, so implicitly depends on the axial and radial position $(z,r)$.

Substituting \eqn{eqn:sol} into the LHS of \eqn{eqn:integral}, and evaluating the integral over $\Delta_{v_z}$ yields 
\begin{equation}
\frac{\partial\,\ln\Omega^2}{\partial z}
=2\,\pi^{3/2}K\sum_{j=0}^2 a_j V_\nu^{(j+1)}(\Delta)\,(\varepsilon\,\Omega)^{2}+O(\Omega^6),\label{eqn:integral2}
\end{equation}
where we have introduced a pseudo-parameter \mbox{$\varepsilon\equiv1$} which we will use later to find a power-series solution, \mbox{$\nu=\DW/\LW$}, and we define a generalised Voigt profile
\begin{equation}
V^{(n)}_\nu(\Delta)=\int_{-\infty}^{\infty} d\Delta_{v_z} \frac{e^{-{(\Delta-\Delta_{v_z})^2}/{\nu^2}}}{\pi^{3/2}(1+{{\Delta_{v_z}^2}})^n}\nonumber%\label{eqn:V2}
\end{equation}
so that the conventional Voigt profile is  
$$V_\nu(\Delta)\equiv V_\nu^{(1)}(\Delta) =\re\{{e^{-{(\Delta+i)^2}/{\nu ^2}} \text{erfc}[{(1-i\Delta)}/{\nu }]}/{\sqrt{\pi }}\}.$$

$V^{(2)}_\nu$ may be evaluated by noting that 
\begin{equation}
\frac{1}{(1+x^2)^2}=\lim_{\xi\rightarrow1}\left\{\frac{\xi^2}{\left(\xi^2-1\right) \left(\xi^2
   x^2+1\right)}+\frac{1}{\left(1-\xi^2\right) \left(x^2+1\right)}\right\}.\nonumber
\end{equation}
Convolving this   sum of two Lorentzians with a Gaussian thus yields a  sum of Voigt functions,
\begin{eqnarray}
V^{(2)}_\nu(\Delta)&=&   \lim_{\xi\rightarrow1}\left\{\frac{\xi^2}{\left(\xi^2-1\right) }V_{\xi\,\nu}(\xi\, \Delta)+\frac{1}{\left(1-\xi^2\right)}V_{\nu}( \Delta)\right\}\nonumber\\
   &=&\textrm{Re}\{\frac{ \nu +\sqrt{\pi } e^{-\frac{(\Delta +i)^2}{\nu ^2}} \left( i \Delta +\nu
   ^2/2-1\right) \erfc[\frac{1-i \Delta }{\nu }]}{ \pi  \nu ^3}\}.\nonumber
\end{eqnarray}
Higher order correction terms can be calculated iteratively, using the same method.  In particular
\begin{widetext}
\begin{equation}
V^{(3)}_\nu(\Delta)=\re\left\{\left({2 \nu\,  (2 i\, \Delta +3\, \nu ^2-2)}+{\sqrt{\pi } e^{-{(\Delta +i)^2}/{\nu ^2}}\text{erfc}\big[{(1-i \Delta })/{\nu }\big] (2 (-2+3 i \,\Delta ) \nu
   ^2-4 (\Delta +i)^2+3 \nu ^4) }\right)/({8 \pi  \nu ^4})\right\}.\nonumber
\end{equation}
\end{widetext}
These are shown in \fig{fig:Voigt}.

Equation (\ref{eqn:integral2}) can be solved analytically for $\Omega^2(z)$ (for any order of expansion), however it  becomes cumbersome.  Instead, we expand in powers of $\varepsilon$, by substituting $\Omega^2(z,r)=\Omega_0^2(z,r)+\varepsilon\,\Omega_1^2(z,r)+\varepsilon^2\Omega_2^2(z,r)+...$ into \eqn{eqn:integral2}, and solving order-by-order in powers of $\varepsilon$, with initial conditions $\Omega^2(0,r)=\Omega_0^2(r)$ (where the radial dependence is parametric).  Physically, this is simply an expansion around the zero-power limit.  We find
\begin{eqnarray}
\Omega_0^2(z,r) &=& \Omega_0^2(r) e^{-p V_\nu(\Delta)},\\
\Omega_1^2(z,r)&=& \Omega_0^4(r) \frac{a_1 e^{-p V_\nu(\Delta)}  (e^{-p V_\nu(\Delta)} -1)V_\nu^{(2)}(\Delta )}{a_0 V_\nu(\Delta )},\\
\Omega_2^2(z,r) &=&\Omega_0^6(r)\left(\frac{a_1^2   e^{-p V_\nu(\Delta  )} (e^{-p
   V_\nu(\Delta  )}-1)^2V_\nu^{(2)}(\Delta  ){}^2}{a_0^2 V_\nu(\Delta  ){}^2}\right.\nonumber\\
   &&\left.\hphantom{\Omega}{}+\frac{a_2 
    e^{-p V_\nu(\Delta  )} (e^{-2 p V_\nu(\Delta  )}-1)V_\nu^{(3)}(\Delta  )}{2 a_0
   V_\nu(\Delta  )}\right),
\end{eqnarray}
where $p=-2\,\pi^{3/2}  K\,a_0 z >0$. 

Since the intensity is proportional to $\Omega^2(z,r)$, we compute the transmission  $z=L$ as 
\begin{equation}
\mathcal{T}=\frac{\int_0^\infty r\,dr\,\Omega^2(L,r)}{\int_0^\infty r\,dr\,\Omega_0^2(r)}.
\end{equation}
This gives Eq.\,(2) of the main text:
 \begin{eqnarray}
\mathcal{T}&=&e^{-p V_\nu(\Delta)}\Big(1+
q_1{(e^{-p V_\nu(\Delta)}-1) V_\nu^{(2)}(\Delta)}/{V_\nu(\Delta )}\nonumber\\
&&{}\hphantom{e^{-p V_\nu()}}+  q_2\big({(e^{-p V_\nu(\Delta)}-1) V_\nu^{(2)}(\Delta)}/{V_\nu(\Delta )}\big)^2\nonumber
   \\
  &&{}\hphantom{e^{-p V_\nu()}} + q_3{(e^{-2p V(\Delta ,\nu )}-1) V_\nu^{(3)}(\Delta)}/{V_\nu(\Delta)}\Big),\nonumber
   \end{eqnarray}
   where $q_1=(a_1\overline{\Omega_0^4})/(a_0\overline{\Omega_0^2})$, $q_2=(a_1^2\overline{\Omega_0^6})/(a_0^2\overline{\Omega_0^2})$ and \mbox{$q_3=(a_2\overline{\Omega_0^6})/(2a_0\overline{\Omega_0^2})$}, and $\overline{\Omega_0^n}=\int_0^\infty r\,dr\,\Omega_0^n(r)$.    
   
   It follows that $q_1$ is proportional to the intensity, whilst $q_2$ and $q_3$ are proportional to the square of the intensity.  This is made explicit if  we assume that the input probe beam has a gaussian transverse profile at $z=0$, $ \Omega_0^2(r)={4 \Omega_0^2 e^{-{2 (r/r_0)^2}}}$, and we find $q_1=2\,a_1\Omega_0^2/a_0$, $q_2=16a_1^2\Omega_0^4/(3a_0^2)$ and $q_3=8a_2\Omega_0^4/(3a_0)$.  In particular, $q_2=4q_1^2/3$, so that $q_2$ is determined by $q_1$.  Since we do not know the parameters $a_{1,2}$, or the optical depth we treat $p$ and $q_i$ as fitting parameters for each resonance.

%\bibliography{bib}
